\documentclass[journal]{IEEEtran}
\usepackage{multirow,url,subfigure}
\usepackage{rotating}

\usepackage{xcolor}

\begin{document}
\title{Improving Trajectory Modelling for DNN-based Speech Synthesis by using Stacked Bottleneck Features and Minimum Generation Error Training}

% previous title: \title{Stacked Bottlenecks and Minimum Trajectory Error Training for DNN-based Speech Synthesis}


\author{Zhizheng~Wu,~\IEEEmembership{Member,~IEEE,}
        and~Simon~King,~\IEEEmembership{Fellow,~IEEE}% <-this % stops a space
\thanks{This work was supported by EPSRC under Programme Grant EP/I031022/1 (Natural Speech Technology).}% <-this % stops a space
\thanks{Z. Wu and S. King are with the Centre for Speech Technology Research (CSTR), University of Edinburgh, UK. Email: \{zhizheng.wu, simon.king\}@ed.ac.uk}% <-this % stops a space
%\thanks{Manuscript received xx. xx, 2015; revised January 11, 2007.}
}


\markboth{Submitted to IEEE/ACM Transactions on Audio, Speech, and Language Processing 2016}%
{Shell \MakeLowercase{\textit{et al.}}: Bare Demo of IEEEtran.cls for Journals}



\maketitle

\begin{abstract}
We propose two novel techniques --- \emph{stacking bottleneck features} and \emph{minimum generation error training criterion} --- to improve the performance of deep neural network (DNN)-based speech synthesis. The techniques address the related issues of \emph{frame-by-frame independence} and \emph{ignorance of the relationship between static and dynamic features}, within current typical DNN-based synthesis frameworks.
Stacking bottleneck features, which are an acoustically--informed linguistic representation, provides an efficient way to include more detailed linguistic context at the input.
The minimum generation error training criterion minimises overall output trajectory error across an utterance, rather than minimising the error per frame independently, and thus takes into account the interaction between static and dynamic features. The two techniques can be easily combined to further improve performance. We present both objective and subjective results that demonstrate the effectiveness of the proposed techniques. The subjective results show that combining the two techniques leads to significantly more natural synthetic speech than from conventional DNN or long short-term memory (LSTM) recurrent neural network (RNN) systems.
\end{abstract}



\section{Introduction}
\label{sec:introduction}

Statistical parametric speech synthesis (SPSS)~\cite{zen2009statistical} has advanced particularly rapidly in the last decade, as seen across the annual Blizzard Challenges~\cite{King2014_loquens}, and can produce highly-intelligible synthesised speech with acceptable naturalness. However, although it offers greater flexibility than the other mainstream technique of unit selection~\cite{hunt1996unit}, the naturalness of speech generated by SPSS is still too low.

There are many factors that underlie this, and acoustic modelling is a key one, as discussed in~\cite{zen2009statistical}. The task of modelling the complex relationship between linguistic representations derived from text input and acoustic features computed from speech waveforms is of course very difficult. In this work, we propose two novel techniques to improve this acoustic modelling. Both techniques target improved modelling of the temporal natural of speech, but in different ways: one via the input linguistic features, the other via the output speech parameters. Separately, each of them results in improvements to the subjective naturalness of the synthesised speech, and their combination gives a further improvement.

\subsection{Related work}
\label{sec:related_work}

Very substantial effort has been devoted to acoustic modelling in the hidden Markov model (HMM) speech synthesis framework. Amongst the many proposed techniques, we highlight just a few of the most influential. In~\cite{wu2006minimum}, a minimum generation error training criterion was proposed to address an inconsistency between training and generation criteria, and the lack of interaction between static and dynamic features during training. In~\cite{zen2007reformulating}, the so-called trajectory HMM was proposed to explicitly model the relationships between static and dynamic features. As a complement to improving the acoustic model itself, enhancement techniques such as global variance~\cite{tomoki2007speech} and modulation spectrum enhancement~\cite{takamichi2014postfilter} aim to mitigate the lack of variation in generated parameter trajectories that results from using an incorrect acoustic model. Although such enhancement techniques do not reduce objective error (e.g., lower spectral distortion w.r.t. a natural speech reference), significant improvements in subjective naturalness are obtained. However, none of the above techniques address what is perhaps the most fundamental problem of HMM-based speech synthesis: across-context averaging via decision tree clustering, which has been identified as a major contributing factor to reduced naturalness~\cite{tom2015attributing}.

More recently, following on from successes in automatic speech recognition~\cite{hinton-2012}, artificial neural networks have re-emerged as acoustic models for SPSS~\cite{ling2015deep}.
By the 1990s, artificial neural networks had already been employed as feature extractors from text input to produce linguistic features~\cite{karaali1998high}, as acoustic models to map linguistic features to vocoder parameters~\cite{weijters1993speech, cawley1993lsp, tuerk1993speech}, and to predict segment durations~\cite{riedi1995neural}. One prominent theme in more recent studies is the use of neural architectures to replace Gaussian mixture models (GMMs) associated with leaf nodes of decision trees, such as the restricted Boltzmann machines (RBMs) in~\cite{ling-2013}, where RBMs were claimed to better learn spectral detail, resulting in better quality synthesised speech. In~\cite{kang2013multi, kang2014statistical}, a deep belief network (DBN) was employed as a deep generative model of the joint probability distribution between linguistic and acoustic features. Other variants on the neural architectures applied to SPSS include the use of deep mixture density networks to predict probability density functions over acoustic features given the corresponding linguistic features~\cite{zen2014deep} and a trajectory real-valued neural autoregressive density estimator to model acoustic parameter trajectories as well as across-feature dependencies~\cite{uria2015modelling}. Neural approaches have been applied to enhancement too, such as the deep generative model in~\cite{chen2015deep} acting as a post-filter to enhance the quality of speech synthesised from an HMM-based system.

The most popular way to use neural networks in SPSS is with a deep feed-forward neural network (DNN) as a conditional model to map linguistic features to vocoder parameters directly~\cite{ze2013statistical, lu2013combining, qian2014training, wu2015deep, watts2016fromhmm}. This can be viewed as replacing the decision tree used in HMM-based speech synthesis with a more powerful regression model~\cite{ze2013statistical,king2015reading}. DNNs have other advantages, including the ability to model high-dimensional acoustic parameters (e.g., the spectrum~\cite{cassia2015perceptual}), and the availability of techniques such as multi-task learning~\cite{wu2015deep,hu2015fusion}. However, a limitation of standard DNN implementations is that the mapping is performed frame by frame without considering contextual constraints, other than those encoded in the input linguistic features. Even though dynamic features are part of the output acoustic feature vector -- because they are needed as a constraint to generate smooth parameter trajectories at synthesis time -- contextual constraints between statics and deltas (or equivalently between successive frames of statics) are not explicitly modelled during training.

One way to model contextual constraints is proposed in~\cite{fan2014rnn}: a bidirectional long short-term memory (LSTM)-based recurrent neural network (RNN) to map a sequence of linguistic features to the corresponding sequence of acoustic features. An LSTM with a recurrent output layer is proposed in~\cite{zen2015unidirectional} to smooth acoustic features across consecutive frames. A systematic investigation on the architectures of gated recurrent neural network can be found in~\cite{wu2016investigating}. These studies formulate speech synthesis as a sequence-to-sequence mapping problem and provide evidence that a better model of speech parameter \emph{trajectories} results in better synthetic speech.

\subsection{Contributions of this work}

We propose some alternative way to include contextual temporal constraints during both training and generation. The proposed framework is easy to train and has an additional benefit that parts of the model can be trained on out-of-domain, lower-quality data (e.g., corpora used in speech recognition). We offer two contributions:

First, we propose \emph{stacked bottleneck features} as a way to include more detailed
linguistic contextual constraints at the input\footnote{Preliminary results were published in~\cite{wu2015deep}. Here, we significantly extend that work with a systematic analysis of the bottleneck features (e.g., the positioning of the bottleneck layer and the contextual width of bottleneck features) and an analysis of the use of out-of-domain data to train the bottleneck feature extractor.}. We train a first network with a bottleneck hidden layer, which has a much smaller number of units than the other hidden layers. The input to this network is the usual set of linguistic features~\cite{tokuda2002hmm}, and the output is some representation of the corresponding speech signal (e.g., the usual vocoder parameters). The activations of the units in the bottleneck layer are thus a lower-dimensional embedding of the input linguistic features that captures information useful for predicting the acoustic features (due to the supervised training) but discards irrelevant or erroneous information (i.e., denoises the features). Then, we stack bottleneck features from multiple consecutive frames around the central frame, and concatenate these stacked bottlenecks with the usual linguistic features. We use these combined features as input to a second neural network to predict vocoder parameters and thus to perform speech synthesis. Note that, because the bottleneck layer size is small (e.g., 32), stacking bottleneck features from multiple consecutive frames does not increase the dimensionality of the input features very much.

Second, we apply a sequential training criterion -- \emph{minimum generation error (MGE)} --- for DNNs\footnote{Preliminary results were published in~\cite{wu2015minimum}. Here we add a comprehensive description of the theory, plus further implementation details and experimental analysis. A similar idea, called sequence generation error, has been independently proposed by Fan et. al.~\cite{fan2015sequence}, and published at the same time as~\cite{wu2015minimum}.}, which is inspired by minimum generation error for HMM-based speech synthesis~\cite{wu2006minimum} and sequence error minimisation for voice conversion~\cite{xie2014sequence}. In a typical conventional implementation of a DNN for speech synthesis, dynamic features (extracted from the sequence of static features) are included as part of the output vector; but, the relationship between the static and dynamic features is neglected during training. The MGE criterion minimises the utterance-level vocoder parameter trajectory error rather than the sum of frame-wise mean squared errors. In this way, the MGE criterion explicitly accounts for the relationship between static and dynamic features and correctly uses the dynamic constraints in the training phase. 

Because the two techniques proposed in this paper are applied at different places in the architecture, namely the linguistic input layer and acoustic output layer, it is possible and natural to combine them. We provide experimental results for this combination.

%There are at least two advantages of the proposed framework: a) the use of additional data to train the bottleneck network. As the bottleneck network is not used to predicted vocoder parameters, it is possible to use low-quality data to train the bottleneck network. Low-quality data is relative easy to obtain. b) Low computational complexity. 

\section{Problem statement}
\label{sec:problem_statement}

To place our proposed methods in context, we briefly review DNN-based speech synthesis and discuss the limitations of typical DNNs as used for speech synthesis, that our proposed methods address.

\subsection{DNN-based speech synthesis}

DNN-based speech synthesis comprises offline training and runtime generation phases. During training, a DNN learns the complex relationship between input linguistic features $\mathbf{x}_{t}$ and corresponding output acoustic features $\mathbf{o}_{t}$ : 
\begin{equation}
\mathbf{o}_{t} = \mathcal{F}(\mathbf{x}_{t}) + \mathbf{e},
\end{equation}
where $\mathcal{F}(\cdot)$ is the mapping function realised by the trained DNN, and $\mathbf{e}$ is the modelling error.
Usually, the acoustic features $\mathbf{o}_{t}$ consist of static features $\mathbf{c}_{t}$, also called vocoder parameters, and corresponding dynamic features $\Delta \mathbf{c}_{t}$ and $\Delta^2 \mathbf{c}_{t}$, written as
\begin{equation}
\mathbf{o}_{t} = [ \mathbf{c}^{\top}_{t}, \Delta \mathbf{c}^{\top}_{t},  \Delta^2 \mathbf{c}^{\top}_{t}]^{\top}.
\end{equation}
The dynamic features are used as a constraint to produce smooth parameter trajectories during generation.
The dynamic features are computed from the sequence of static features. Hence, a sequence of observed acoustic features $\mathbf{O}=[\mathbf{o}_{1}^{\top}, \mathbf{o}_{2}^{\top}, \cdots, \mathbf{o}_{T}^{\top}]^{\top}$ can be calculated from a sequence of static features $\mathbf{C}=[\mathbf{c}_{1}^{\top}, \mathbf{c}_{2}^{\top}, \cdots, \mathbf{c}_{T}^{\top}]^{\top}$ by 
\begin{equation}
\mathbf{O} = \mathbf{W} \mathbf{C},
\end{equation}
where $\mathbf{W}$ is a matrix that contains the coefficients used to compute static, delta and delta-delta features from a sequence of static features $\mathbf{C}$. Details can be found in~\cite{tokuda2000speech}. 

To train a DNN, the usual objective is to minimise, in a frame-wise fashion, the error $D(\hat{\mathbf{o}_{t}}, \mathbf{o}_{t})$ between predicted  $\hat{\mathbf{o}_{t}}$ and observed acoustic features $\mathbf{o}_{t}$; this objective function can be written as
\begin{equation}
\label{eq:dnn_obj}
D(\hat{\mathbf{o}_{t}}, \mathbf{o}_{t}) = (\hat{\mathbf{o}_{t}} - \mathbf{o}_{t})^{\top}  (\hat{\mathbf{o}_{t}} - \mathbf{o}_{t}).
\end{equation}
where $D(\hat{\mathbf{o}_{t}}, \mathbf{o}_{t})$ is frame-wise error computed at the output. To minimise this error, the classic gradient descent algorithm \emph{back-propagation}~\cite{rumelhart1986learning} is typically used. The gradients of the DNN parameters can be calculated by taking derivatives of $D(\hat{\mathbf{o}_{t}}, \mathbf{o}_{t})$ with respect to model parameters $\mathbf{\lambda}$ :
\begin{eqnarray}
\label{eq:gradient}
\frac{\partial D(\hat{\mathbf{o}_{t}}, \mathbf{o}_{t})}{\partial \mathbf{\lambda}} = \frac{\partial D(\hat{\mathbf{o}_{t}}, \mathbf{o}_{t})}{\partial \mathbf{o}_{t}} \frac{\partial \mathbf{o}_{t}}{\partial \mathbf{\lambda}},
\end{eqnarray}
where
\begin{eqnarray}
 \label{eq:bp_error}
 \frac{\partial D(\hat{\mathbf{o}_{t}}, \mathbf{o}_{t})}{\partial \mathbf{o}_{t}}  = \hat{\mathbf{o}_{t}} - \mathbf{o}_{t}
\end{eqnarray}
is the error to be back-propagated through the network from the top output layer to the bottom input layer, and the gradients of model parameters at each layer can be calculated through this back-propagation process. In practice, a mini-batch gradient descent method is usually applied, for faster convergence and more stable behaviour~\cite{hinton2010rbm}; this is possible because the error at each frame can be calculated independently.

At generation time, given a sequence of linguistic features $\mathbf{X}$, the corresponding acoustic features $\hat{\mathbf{O}}$ can be generated from the trained DNN by performing a forward propagation once per frame. To generate smooth parameter trajectories, the maximum likelihood parameter generation (MLPG) algorithm~\cite{tokuda2000speech} is used, to take the dynamic feature constraints into account. Recall that the DNN predicts both static and dynamic features (although without ensuring consistency between them). The MLPG algorithm can be expressed as
\begin{equation}
\label{eq:mlpg}
\hat{\mathbf{C}} = (\mathbf{W}^{\top} \mathbf{U}^{-1} \mathbf{W})^{-1} \mathbf{W}^{\top} \mathbf{U}^{-1} \hat{\mathbf{O}},
\end{equation}
where $\hat{\mathbf{C}}$ is the predicted static acoustic feature sequence (i.e., trajectory), which will be used to reconstruct the speech waveform, and $\mathbf{U}$ is the covariance matrix, which is computed from the training data in DNN-based framework. Using MLPG is important for good quality synthesised speech~\cite{chen2010perceptual, hashimoto2015effect}.

\subsection{Limitations}

Although DNNs have been reported to achieve significant improvements over HMMs for speech synthesis, as we reviewed in Section~\ref{sec:related_work}, there are at least two limitations in current DNN implementations:

\subsubsection{Frame-by-frame independence} Each frame's acoustic features are predicted from that frame's linguistic features without any contextual constraints other than those encoded in the linguistic features. Acoustic context, which is so important in speech, is not explicitly considered either during training or in the forward propagation step of generation.

\subsubsection{Neglecting the relationship between static and dynamic features} During the generation process, dynamic features are used by the MLPG algorithm to generate smooth parameter trajectories. But it relies on \emph{potentially inconsistent} static and dynamic features predicted by the DNN. It should be beneficial to include the dynamic feature constraints during training.

The two techniques that we propose for addressing these limitations are now introduced in Sections ~\ref{sec:bottleneck} and~\ref{sec:mte_criterion}.

\section{Stacked bottleneck features}
\label{sec:bottleneck}

A straightforward approach might be to stack linguistic input features from several consecutive frames at the input. These linguistic features are usually extracted per phone and they include more slowly-changing word and phrase level information; then,  information from forced-alignment is used to interpolate the phoneme-level linguistic features to obtain a frame-level input for the DNN. Therefore, stacking multiple frames of linguistic features would result in high-dimensional, sparse and highly-redundant features, which still may not be effective in capturing contextual constraints. 

So, we propose instead to stack acoustically-informed bottleneck features, which are intended to capture all the relevant information from the input linguistic representation. Bottleneck features are the activations at a bottleneck layer in a DNN. This layer has a relatively small number of hidden units compared to the other hidden layers in the same network. Bottleneck features have been extensively employed in automatic speech recognition (ASR) as a compact representation of acoustic features~\cite{grezl2008optimizing, yu2011improved, sainath2012auto}. For speech synthesis, bottleneck features can be viewed as a compressive transform of the linguistic features, extracted at the frame level. Because the network, in which the bottleneck layer is situated, is trained in a supervised fashion using acoustic features (which of course change every frame, reflecting the continuous nature of speech signals) the bottleneck features can also capture fine-grained sub-phonetic temporal variations.

%Fig.~\ref{fig:bottleneck} illustrates an example sequence of bottleneck features for one utterance. It shows that bottleneck features do indeed capture something of the continuous nature of speech signals; they are not piecewise constant within each phone, for example. 
%\begin{figure}[!htb]
%\centering
%\includegraphics[scale=0.5]{figs/bottleneck.pdf} 
%\caption{An example sequence of bottleneck features for one utterance (200 frames per second).}
%\label{fig:bottleneck}
%\end{figure}
%

\begin{figure}[!htb]
\centering
\includegraphics[scale=0.45]{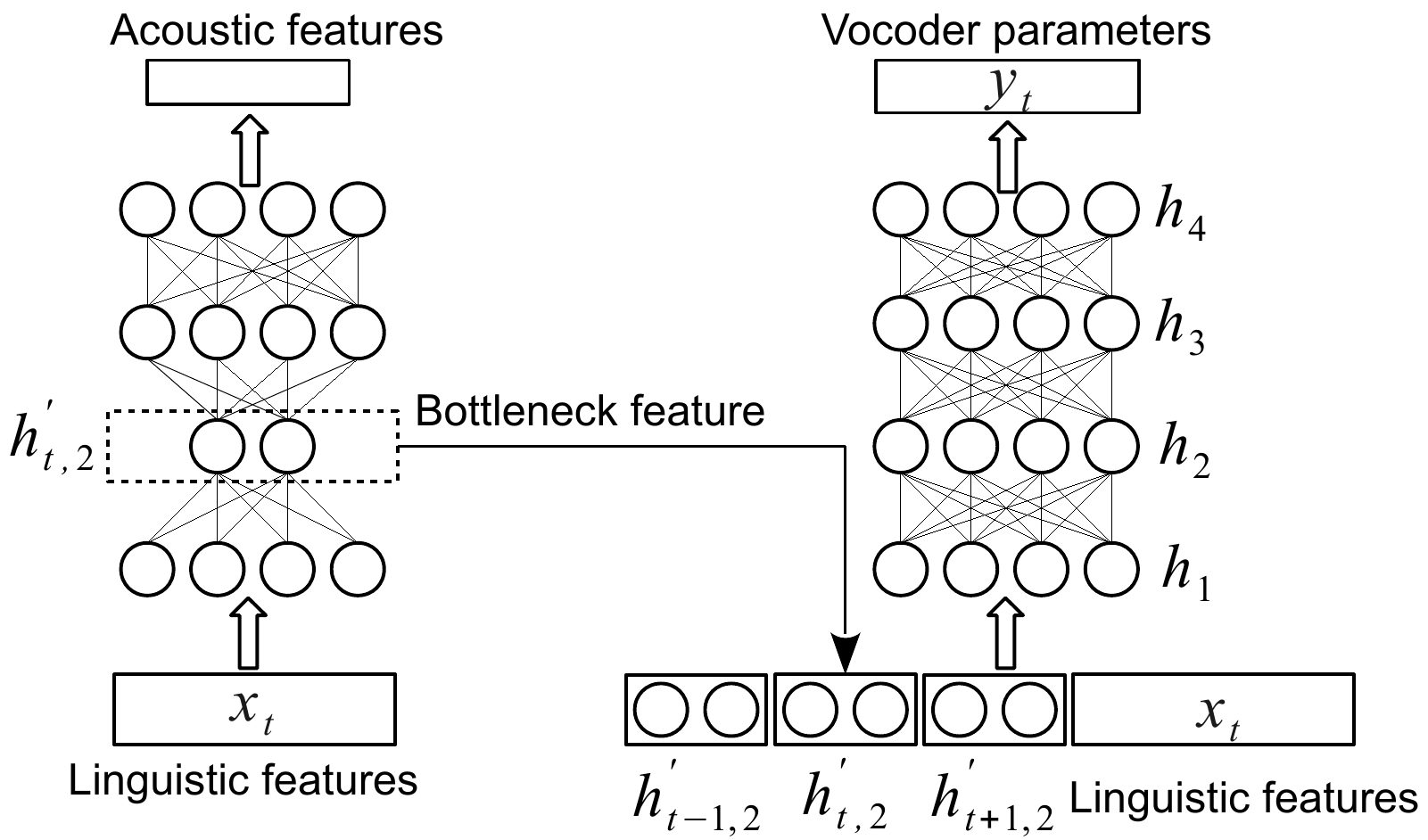} 
\vspace{-3mm}
\caption{On the left is a deep neural network (DNN) with a bottleneck layer. In this example, the bottleneck features for three consecutive frames are stacked as input to the network on the right, which is the synthesis network. In practice, more than three frames can be included. $\mathbf{h}_{t, 2}^{'}$ is the vector of bottleneck features for frame $t$.}
\label{fig:dnn_bottleneck}
\end{figure}

Fig.~\ref{fig:dnn_bottleneck} illustrates the architecture of a DNN system that employs stacked bottleneck features. The left-hand network is a bottleneck network with four hidden layers of which the second layer is the bottleneck layer\footnote{The numbers of layers in the example networks are only for illustration. We used a different setting in the experiments detailed in Section~\ref{sec:exp}.}. The left-hand network is used to extract bottleneck features, which are then stacked as input to the right-hand network. Since the dimensionality of the bottleneck features is small (e.g., 32), stacking such features from multiple frames does not increase input dimensionality much, nor does it increese the computational complexity of the synthesis network (the right-hand network in Fig.~\ref{fig:dnn_bottleneck}). The method proceeds as follows:
\begin{itemize}
	\item[(a)] Train a network with a bottleneck layer. The input comprises the linguistic features and the output is the corresponding acoustic features;
	\item[(b)] Given a sequence of linguistic features, perform a forward propagation through the bottleneck network to generate bottleneck features, frame by frame;
	\item[(c)] Stack these bottleneck features from several consecutive frames around the current frame alongside the linguistic features;
	\item[(d)] With linguistic features and stacked bottlenecks as input, train a synthesis network to predict vocoder parameters;
	\item[(e)] To perform synthesis from a sequence of linguistic features, follow steps (b) and (c) to obtain input features, then make a forward pass through the synthesis network to generate vocoder parameters, and thence synthetic speech.
\end{itemize}
It is important to note that the bottleneck network is never used to generate synthetic speech. So, the output of the bottleneck network may be any kind of acoustic feature (e.g., Mel-Frequency Cepstral Coefficients (MFCCs)). Neither does the bottleneck network have to be trained on the same data as the synthesis network; for example, it is possible to use additional data from other speakers to train the bottleneck network. We will investigate the performance of various such system configurations in Section~\ref{sec:exp}.

\section{Minimum generation error training criterion}
\label{sec:mte_criterion}

In the previous section, we introduced the idea of stacking bottleneck features to explicitly include contextual constraints at the input to a DNN. We now turn to the output features, where the interaction between the static and dynamic features at the output is still neglected in conventional DNN training. The second contribution of this paper is a novel training criterion: minimum generation error (MGE). The MGE criterion minimises the utterance-level trajectory error rather than frame-by-frame error. This is because it explicitly accounts for the interaction between the static and dynamic features, whereas conventional DNNs for SPSS treat the dynamic features no differently to static features. 

The basic idea of the MGE criterion is to minimise the error of the output after MLPG; i.e., the vocoder parameter trajectories which will actually be used to reconstruct the speech waveform, rather than to minimise the error of the input to MLPG, namely the static and dynamic acoustic features. We define the trajectory error as the Euclidean distance between the predicted $\hat{\mathbf{C}}$ and reference $\mathbf{C}$ static parameter trajectories, and the new objective function is defined as 
\begin{eqnarray}
D(\hat{\mathbf{C}}, \mathbf{C}) & = & (\hat{\mathbf{C}} - \mathbf{C})^{\top} (\hat{\mathbf{C}} - \mathbf{C}) \\
             				    & = & (\mathbf{R}\hat{\mathbf{O}} - \mathbf{C})^{\top} (\mathbf{R}\hat{\mathbf{O}} - \mathbf{C}),
\end{eqnarray}
where $$\mathbf{R} = (\mathbf{W}^{\top} \mathbf{U}^{-1} \mathbf{W})^{-1} \mathbf{W}^{\top} \mathbf{U}^{-1}$$
is the matrix that performs MLPG, given static and dynamic features, similar to that in Eq. (\ref{eq:mlpg}). In practice, mean-variance normalisation is performed to $\hat{\mathbf{C}}$ and $\mathbf{C}$ for trajectory error calculation. The mean and variance are pre-calculated from the training data.

Compared to Eq. (\ref{eq:dnn_obj}), the new objective function $D(\hat{\mathbf{C}}, \mathbf{C})$ is calculated from the smoothed trajectory (after MLPG) rather that the direct output of the DNN.

As with any conventionally-trained DNN, a gradient descent algorithm can be employed to train the network. With the new error function, the gradients of the DNN model parameters can be calculated as
\begin{eqnarray}
\frac{\partial D(\hat{\mathbf{C}}, \mathbf{C})}{\partial \mathbf{\lambda}} & = & \frac{\partial D(\hat{\mathbf{C}}, \mathbf{C})}{\partial \hat{\mathbf{O}}} \frac{\partial \hat{\mathbf{O}}}{\partial \mathbf{\lambda}} \\
& = & \frac{\partial D(\mathbf{R}\hat{\mathbf{O}}, \mathbf{C})}{\partial \hat{\mathbf{O}}} \frac{\partial \hat{\mathbf{O}}}{\partial \mathbf{\lambda}},
\end{eqnarray}
where 
\begin{eqnarray}
 \label{eq:mte_error}
 \frac{\partial D(\mathbf{R}\hat{\mathbf{O}}, \mathbf{C})}{\partial \hat{\mathbf{O}}}  =  (\hat{\mathbf{C}} - \mathbf{C})^{\top} \mathbf{R}.
\end{eqnarray}
The only difference between the new criterion and the conventional frame-wise mean squared error criterion is the way in which the output errors to be back-propagated through the network are calculated. The difference can be seen in Eq. (\ref{eq:mte_error}) and Eq. (\ref{eq:bp_error}). The method for computing the gradients at lower layers is unchanged, that is to compute the gradients through $\frac{\partial \hat{\mathbf{O}}}{\partial \mathbf{\lambda}}$. Similar to Eq. (\ref{eq:gradient}), only $\frac{\partial \hat{\mathbf{O}}}{\partial \mathbf{\lambda}}$ is directly related to the model parameters. 

Performing back-propagation with the new criterion involves the following steps:
\begin{itemize}
%	\item[(a)] Perform mean-variance normalisation to the acoustic features, and perform min-max normalisation to the linguistic features;
	\item[(a)] Initialise the weights for the MGE-DNN from a conventionally-trained DNN (i.e., using MMSE);
	\item[(b)] Given a sequence of input linguistic features, perform a forward propagation step just as  in conventional training, to predict observation $\hat{\mathbf{O}}$;
	\item[(c)] Restore the mean and variance for $\hat{\mathbf{O}}$ (because mean-variance normalisation is performance for the acoustic features before training);
	\item[(d)] Perform MLPG to generate acoustic feature trajectories $\hat{\mathbf{C}}$ using Eq.(\ref{eq:mlpg});
	\item[(e)] Perform mean-variance normalisation to the predicted trajectories $\hat{\mathbf{C}}$ and the reference trajectories $\hat{\mathbf{C}}$;
	\item[(f)] Calculate the trajectory error using Eq. (\ref{eq:mte_error}) with the mean-variance normalised trajectories;
	\item[(g)] Perform backward propagation just as in the conventional algorithm, except using the error calculated at step (f).
\end{itemize}
In practice, steps (d) to (f), corresponding to Eq. (\ref{eq:mte_error}), are performed dimension by dimension. After the errors of all the dimensions are calculated, we perform step (g). The gradient update process is the same as in the conventional training algorithm.

In conventional training, it is usual to employ mini-batches, in which each mini-batch contains a fixed number of frames from a randomly shuffled version of the training data. With the new criterion, we need to keep  trajectories intact, and therefore all frames from each utterance much be in the same mini-batch, in the original order. We use individual utterances as the mini-batches, so the sizes of mini-batches vary. For synthesis, we proceed exactly as with a conventionally-trained DNN.

In our implementation, most of the computational cost arises from the calculation of $(\mathbf{W}^{\top} \mathbf{U}^{-1} \mathbf{W})^{-1}$. As $\mathbf{U}$ is diagonal, $\mathbf{W}^{\top} \mathbf{U}^{-1} \mathbf{W}$ becomes a banded matrix, and the computational costs can be reduced considerably; we used the bandmat Python library\footnote{\url{https://pypi.python.org/pypi/bandmat/0.5}} to perform inversion of this banded matrix. 

\section{Experiments}
\label{sec:exp}

\subsection{Experimental setups}
\label{sec:setups}

We conducted experiments using a corpus recorded from a British male professional speaker, divided into three subsets: 2400 utterances as training set, 70 utterances as development set, and 72 utterances as testing set. The waveform sampling rate of the corpus is 48 kHz. We used the STRAIGHT vocoder~\cite{kawahara1999restructuring} to extract vocoder parameters --- 60-dimensional Mel-Cepstral Coefficients (MCCs), 25 band aperiodicities (BAPs), and log-scale fundamental frequency ($\log F_{0}$ at a 5ms frame step, and we employed the same vocoder to reconstruct speech waveforms during synthesis.

As reported in our previous work~\cite{wu2015deep} and other previous studies~\cite{qian2014training,hashimoto2015effect}, DNN-based systems are generally significantly better than HMM-based ones, and therefore we only included DNN and LSTM baselines, and no HMM systems. We also implemented intermediate methods incorporating individual techniques of the proposed framework to examine the effectiveness of each of them, as well as their combination. The systems implemented and compared were: 
\begin{itemize}
	\item \textbf{DNN:} a baseline system based a normal feed-forward deep neural network trained using the conventional frame-by-frame minimum mean squared error (MMSE) criterion. The network has six  hidden layers, each of which has 1024 hidden units. The hyperbolic Tangent activation functions are employed in the lower layers, and a linear activation function at the output layer.

	\item \textbf{LSTM:} a second baseline system based on a long short-term memory (LSTM) network with three feed-forward lower hidden layers each of 1024 units with tangent activation functions (intended to extract features, as suggested in~\cite{fan2014rnn}) plus one LSTM layer with 768 units on top of these feed-forward layers, and finally a linear regression output layer.

	\item \textbf{BN-DNN:} similar to the DNN system, and also trained with frame-wise MMSE, but using stacked bottleneck features and linguistic features as in Fig.~\ref{fig:dnn_bottleneck}. The same vocoder parameters as for the BN-DNN system were used as the acoustic output of the bottleneck network. The architecture of the synthesis network was the same as the DNN system.
	
	\item \textbf{BN-DNN-VB:} same as BN-DNN system, except using a different database (the voice bank database~\cite{veaux2013voice}) to train the bottleneck network.
	
	\item \textbf{BN-DNN-MFC:} same as BN-DNN-VB, except the output features of the bottleneck network were 21-dimensional Mel-Frequency Cepstral Coefficients (MFCCs) and their delta, delta-delta features, in total 63-D, extracted from waveforms that had been downsampled to 16 kHz.
	
	\item \textbf{BN-DNN-WSJ:} same as BN-DNN-MFC, except using the Wall Street Journal (WSJ0+WSJ1) database~\cite{paul1992design} to train the bottleneck network.

	\item \textbf{MGE-DNN:} same as the DNN system, but now employing the proposed minimum generation error (MGE) training criterion. The model parameters were initialised from the fully-trained DNN system above.
	
	\item \textbf{MGE-BN-DNN:} same as the BN-DNN system, but now employing the MGE training criterion. The model parameters were initialised from the fully-trained BN-DNN system. Note that the MGE training is only applied to the synthesis network; the bottleneck network is simply taken from the BN-DNN system.
\end{itemize}

All the systems described above employed the same front-end to extract input linguistic features, which comprised 592 binary and 9 numerical features. The binary features were derived from the linguistic features such as quinphone identities, part-of-speech (POS), and positional information of phoneme, syllable, word and phrase. The input features were normalised to the range $[0.01 \;\; 0.99]$.

For the output vocoder parameters, $F_{0}$ was linearly interpolated before modelling, and a binary feature was used to record the voiced/unvoiced information for each frame. Delta and delta-delta features were calculated for MCCs, BAPs and $F_{0}$. In total, there were 259 features in the output. We applied mean-variance normalisation to the output acoustic features such that they had zero mean and unit variance across the training set. Similar normalisation was also employed to the MFCCs.

The hyper-parameters (i.e. the number of hidden layers, the number of hidden units, learning rate, momentum) of all neural networks were tuned on the development set. For all the systems except LSTM, our implementation employed the CUDAMat library\footnote{\url{https://github.com/cudamat/cudamat}}, which is a Python module for matrix calculations on GPU using CUDA, while for LSTM, we employed the Theano library\footnote{Theano version 0.7: \url{http://deeplearning.net/software/theano/}}.

\subsection{Objective evaluation}

We employed objective measures to tune the systems. Although these objective measures might not always be well correlated with human perception, they provide a practical and effective way to optimise the systems, especially for tuning hyper-parameters. We employed four measures:
\begin{itemize}
	\item \textbf{MCD:} Mel-Cepstral Distortion (MCD) to measure MCC prediction performance.
	\item \textbf{BAP:} a distortion measure for BAPs.
	\item \textbf{$F_{0}$ RMSE:} Root Mean Squared Error (RMSE) to measure $F_{0}$ prediction performance. We note that $F_{0}$ was modelled on a \emph{log-scale}, but the error was calculated on a \emph{linear-scale}.
	\item \textbf{V/UV:} to measure voiced/unvoiced error.
\end{itemize}
For all objective measures, a lower value indicates better performance.

\subsubsection{Effect of the position of the bottleneck layer}

We started with experiments to examine the effects of the position of the bottleneck layer for the BN-DNN system. We fixed the context size (number of bottleneck frames that are stacked for input to the synthesis network) to 9, and varied the placement of the bottleneck layer from the bottom hidden layer to the top hidden layer. Examining all combinations of context size and bottleneck layer placement would involve too many experiments to be practical.

\begin{figure}[!t]
\centering
\includegraphics[scale=0.30]{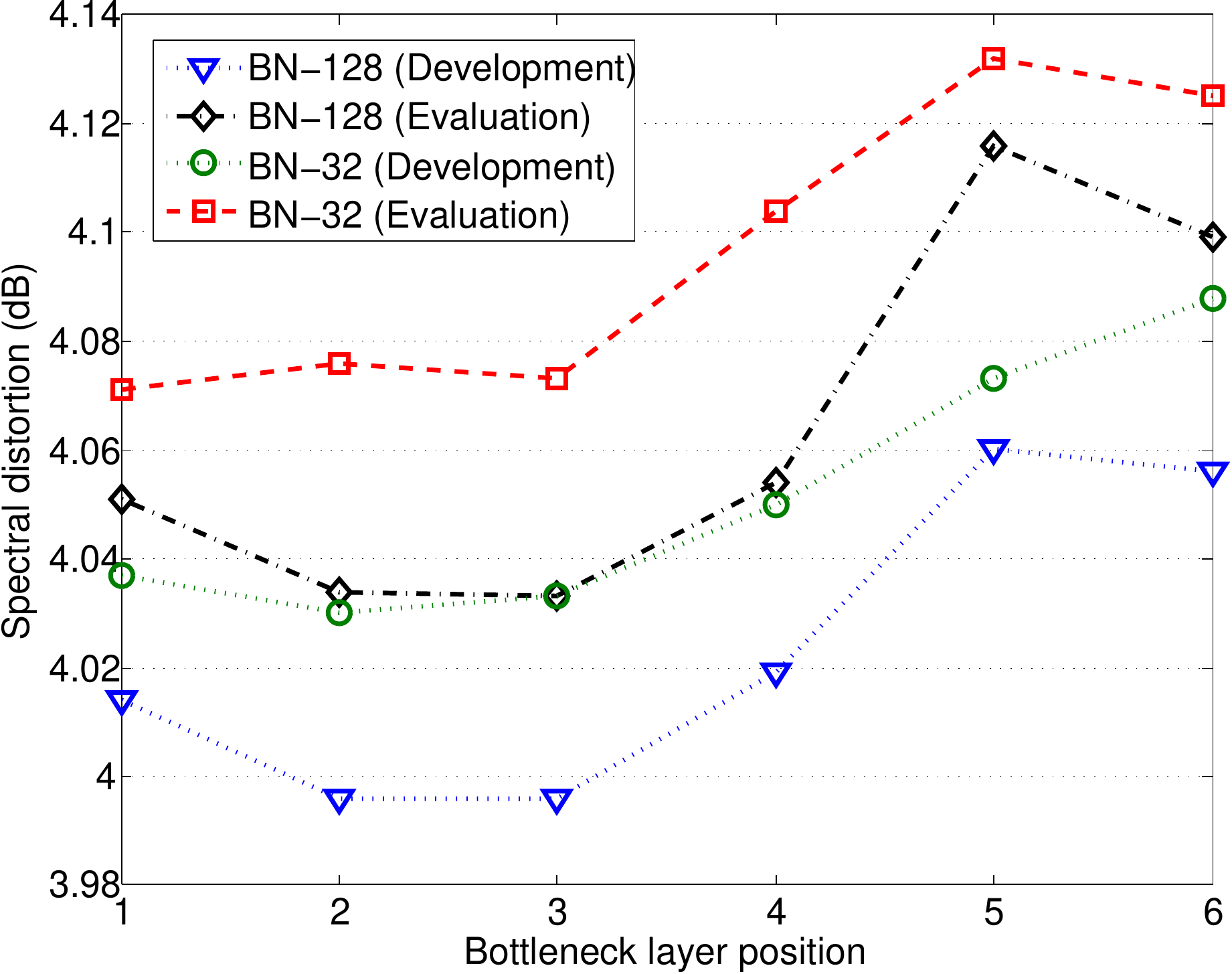} 
\vspace{-3mm}
\caption{Mel-Cepstral Distortion (MCD) as a function of the position of the bottleneck layer. BN-128 and BN-32 indicate bottleneck layer sizes of 128 and 32, respectively. MCD is measured at the output of the synthesis network.}
\label{fig:bottleneck_position}
\end{figure}

When the bottleneck layer is close to the input layer, the bottleneck features are presumed to represent something more akin to the linguistic features than the acoustic features, and vice versa. The objective measure MCD, measured at the output of the synthesis network, is plotted as a function of  bottleneck layer position in Fig.~\ref{fig:bottleneck_position} and shows that lower (closer to the input) positioning of the bottleneck results in lower distortion. In particular, placing the bottleneck layer as the second or third hidden layer works best. BAP distortions also showed a similar pattern, whereas the other objective measures did not substantially vary with bottleneck layer position. We hence place the bottleneck at the second hidden layer in all remaining experiments. We consider the bottleneck layer activations to be a non-linear compression of the linguistic input features, with the compressive transform learned in a supervised way to minimise an acoustic distortion.

\subsubsection{Effect of the contextual size of stacked bottlenecks}

\begin{figure}[!t]
\centering
\includegraphics[scale=0.30]{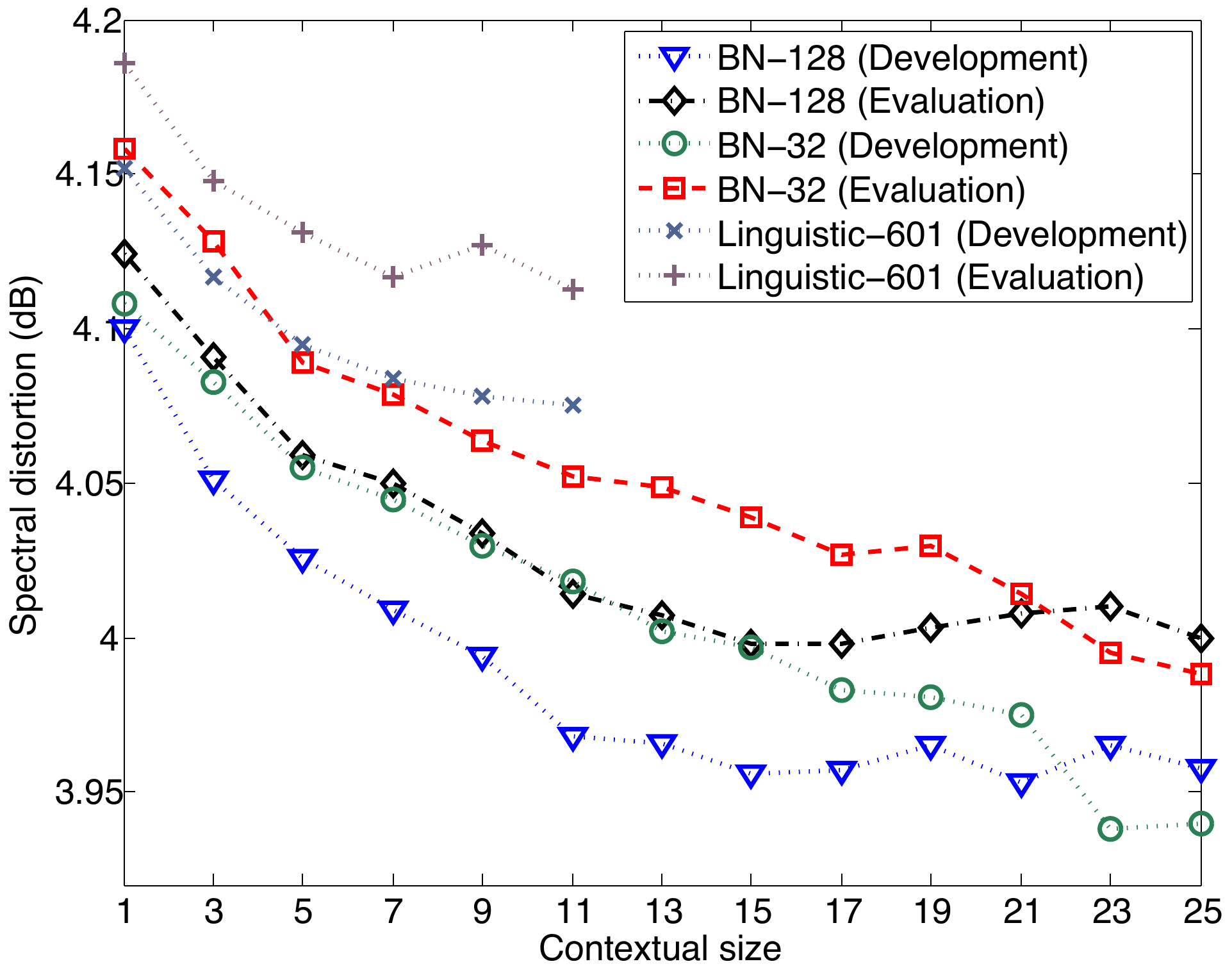} 
\vspace{-3mm}
\caption{Mel-Cepstral Distortion (MCD) as a function of the contextual size of stacked bottlenecks. Linguistic-601 means stacking the 601-dimensional input features directly. MCD is measured at the output of the synthesis network.}
\label{fig:stacking_bottleneck_context}
\end{figure}

We next conducted experiments to examine the effect of the contextual size of the stacked bottleneck features being presented at the input of the synthesis network, again for the BN-DNN system. With the bottleneck always at the second hidden layer of the first network, we varied the contextual size from 1 to 25 and measured MCD at the output of the synthesis network, which is plotted in Fig.~\ref{fig:stacking_bottleneck_context}. In this figure, we also show the effect of stacking up multiple frames of linguistic features at the input to the synthesis network (there are no bottleneck features in this case). The MCD does initially fall, as expected, but not as quickly as when stacking bottleneck features, and plateauing out after a contextual size of around 7 frames. Of course, the dimensionality of the stacked linguistic features becomes very high. For example, for a contextual size of 11: $601 \times 11 = 6611$.

For a bottleneck layer size of 128, the MCD keeps falling until a contextual size of about 15. For a bottleneck layer size of 32, the MCD continues falling until a contextual size of 23; in this case the dimensionality of the stacked bottleneck features is $32 \times 23= 736$ -- much smaller than that of stacked linguistic features, and smaller than for 128-dimensional bottleneck features stacked to a contextual size of 15 ($128 \times 15 = 1920$).

\begin{figure}[!t]
\centering
\includegraphics[scale=0.30]{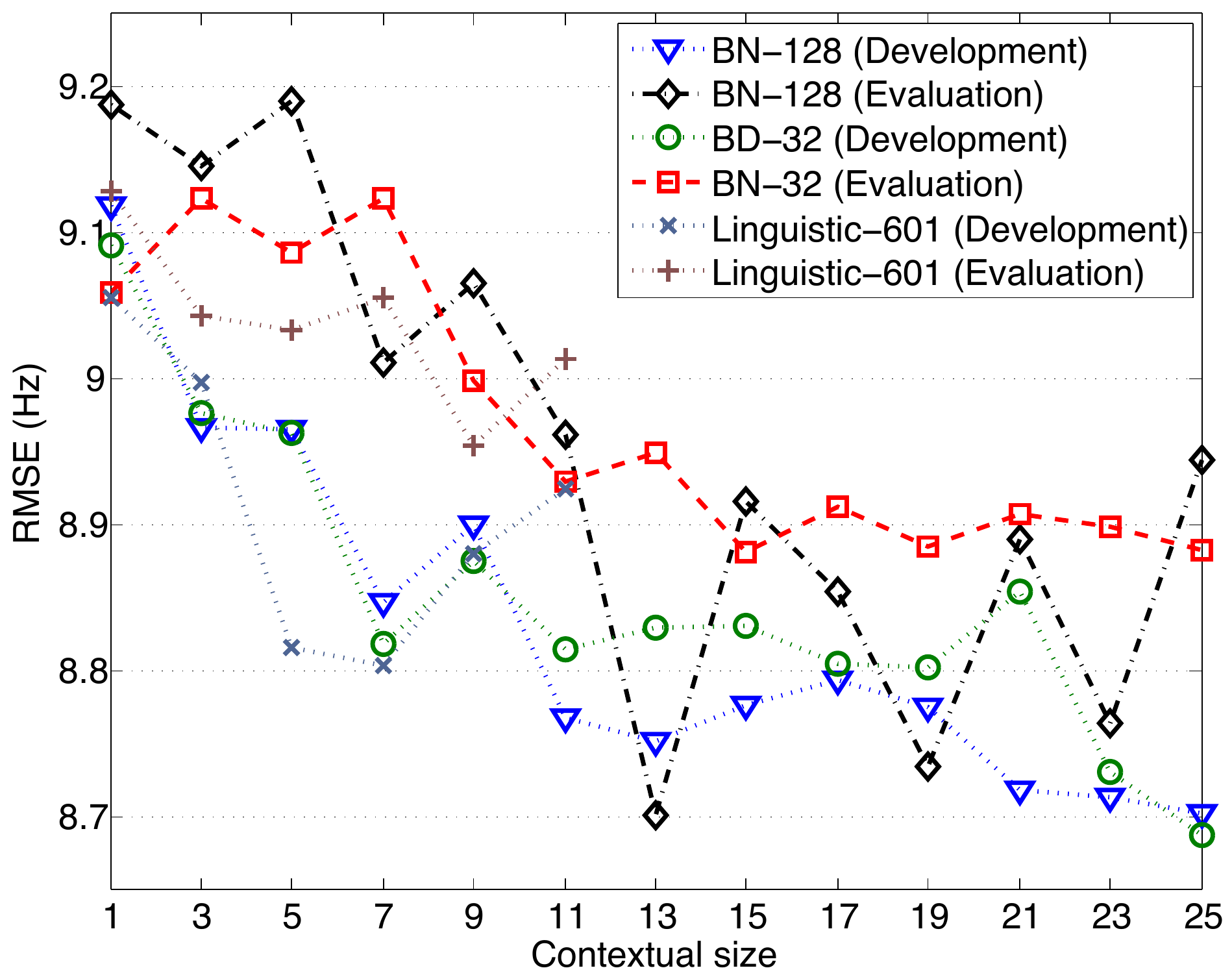} 
\vspace{-3mm}
\caption{$F_{0}$ RMSE $F_{0}$ as a function of the contextual size of stacked bottlenecks. RMSE $F_{0}$ is measured at the output of the synthesis network.}
\label{fig:stacking_bottleneck_context_f0}
\end{figure}

An equivalent plot of RMSE $F_{0}$ is presented in Fig.~\ref{fig:stacking_bottleneck_context_f0}. The behaviour is largely similar to that of MCC, and again a low error is obtained using 32-dimensional bottleneck features stacked to a contextual size of 23.

In summary, a highly effective way to include richer linguistic context is by stacking bottleneck features. A relatively small bottleneck size of 32 provides good performance when stacking 23 contextual frames; this is the configuration used in all remaining experiments.

\subsubsection{Effect of training bottleneck network using out-of-domain data}

As previously noted, the bottleneck layer is providing an acoustically-supervised compression of the linguistic input features. It is interesting to know whether this acoustic supervision has to come from exactly the same data as will later be used to train the synthesis network, and whether the data have to be parametrised in exactly the same way.

We therefore conducted experiments to assess the effects of using out-of-domain data (i.e., different to the main single-speaker British English dataset used to train the synthesis network) to train the bottleneck network. 

We considered three settings, enumerated in Section~\ref{sec:setups}: BN-DNN-VB, BN-DNN-MFC, BN-DNN-WSJ. The BN-DNN-VB and BN-DNN-MFC systems both used the same out-of-domain speech dataset that has been recorded in a high-quality studio for speech synthesis purposes, albeit from non-professional speakers (voice bank), but parametrised differently. BN-DNN-VB used vocoder features extracted from 48 kHz waveforms whilst BN-DNN-MFC used MFCC extracted from 16 kHz waveforms. BN-DNN-WSJ used a database designed for speech recognition and containing American English accented speech. In the voice bank data, there are 96 speakers (41 male, 55 female), each saying about 300 utterances. In total there are 36800 utterances for training the bottleneck network. In the WSJ corpus, there are 283 speakers and a total of about 37000 utterances for training the bottleneck network.

Objective results for the three systems are presented in Table~\ref{tab:summary_objective}. Compared to BN-DNN (which does not use out-of-domain data), the three systems all reduce all four objective measures, except that BN-DNN-VB slightly increases V/UV error. Comparing BN-DNN-VB with BN-DNN-MFC, we see that using data at a lower-sampling rate and simpler acoustic features (MFCCs instead of vocoder parameters) has no effect on the objective measures. Even when using a speech recognition database, containing speakers of a difference accent, to train the bottleneck network (BN-DNN-WSJ), we still get lower distortions than for BN-DNN.

In summary, these objective results demonstrate that using a relatively large amount of out-of-domain speech data to train the bottleneck network improves the synthesis performance compared to only using the smaller single-speaker synthesis data. The sampling rate (and therefore bandwidth) and parametrisation of the speech data has little effect.

\begin{table}[!htb]
\footnotesize
 \scriptsize
\caption{\label{tab:summary_objective} Objective errors for all systems on the evaluation set. MCD and BAP are distortion measures for Mel-Cepstral Coefficients and Band Aperiodicities. RMSE indicates Root Mean Squared Error. V/UV indicates voiced/unvoiced  error.}
\vspace{-5mm}
\begin{center}
    \begin{tabular}{|l|l|l|l|l|}
        \hline
                            & MCD (dB)       & BAP (dB)       & $F_{0}$ RMSE (Hz)    & V/UV (\%)      \\
        \hline\hline
        DNN                 & 4.19      & 1.95      & 9.13        & 4.24      \\
        LSTM                & 4.05      & 1.94      & 8.76        & 3.97      \\
        \hline \hline
        BN-DNN              & 4.00      & 1.92      & 8.90        & 3.97      \\
        BN-DNN-VB           & 3.98      & 1.92      & \textbf{8.57}        & 4.02      \\
        BN-DNN-MFC          & 3.97      & \textbf{1.91}      & 8.61        & \textbf{3.90}      \\
        BN-DNN-WSJ          & 3.97      & 1.92      & 8.59        & 3.96      \\
        \hline \hline
        MGE-DNN             & 4.12      & 1.95      & 8.93        & 4.28      \\
        \textbf{MGE-BN-DNN} & \textbf{3.97}      & 1.92      & 8.89        & 3.96      \\
        \hline
    \end{tabular}
\end{center}
\end{table}

\subsubsection{Effectiveness of minimum generation error training criterion}

\begin{figure}[!t]
\centering
\includegraphics[scale=0.28]{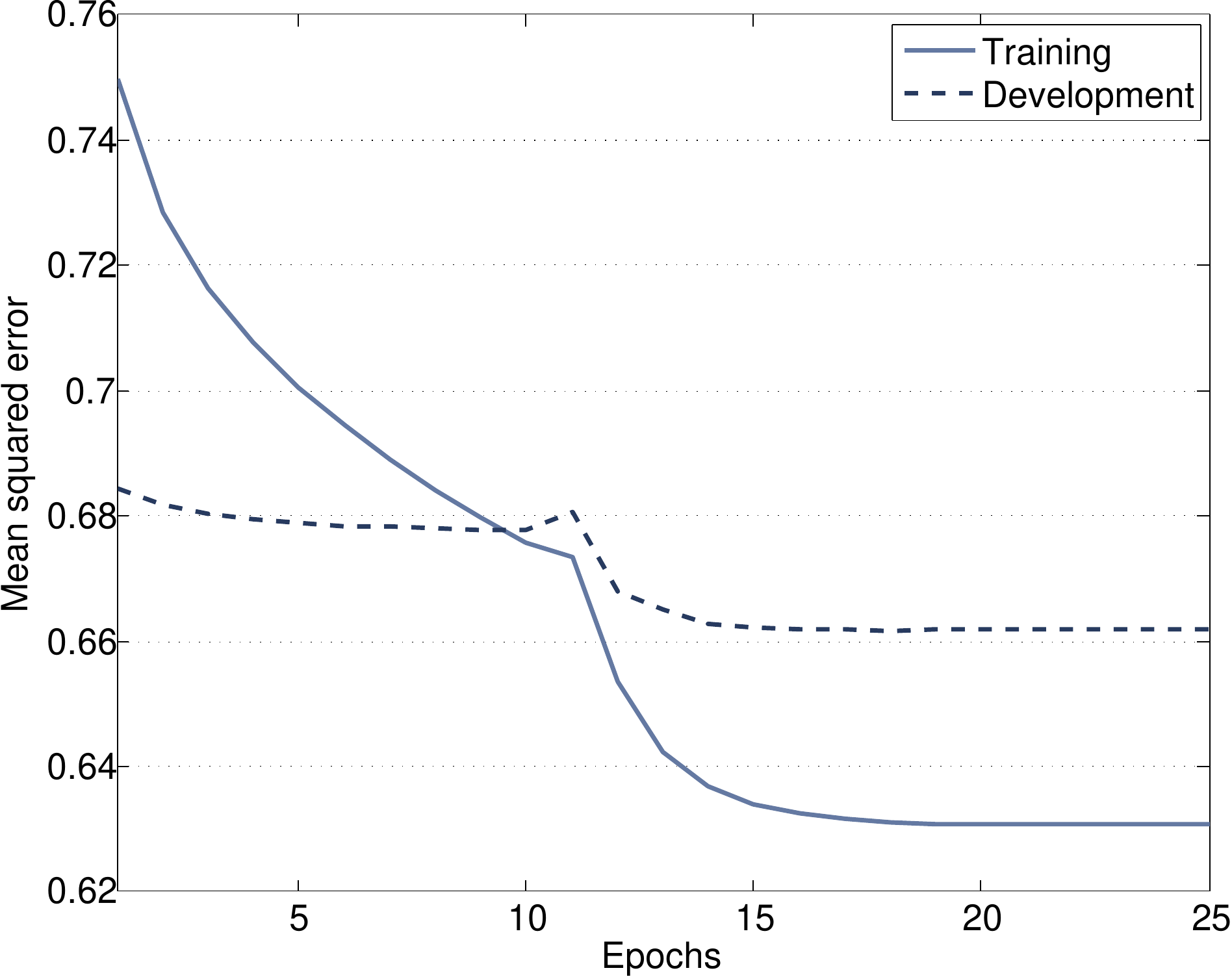} 
\vspace{-3mm}
\caption{Convergence of the proposed minimum generation error training criterion for the DNN system (MGE-DNN). The plot shows total objective error against training epochs.}
\label{fig:mte_convergence}
\end{figure}

The second contribution of this paper is the novel minimum generation error (MGE) training criterion of Section~\ref{sec:mte_criterion}, operating on the output features of the synthesis network. We performed initial experiments to examine the convergence property of training under the MGE criterion, using the objective measure of mean squared error between predicted and reference vocoder parameter trajectories. The error is calculated after MLPG, because the vocoder parameter trajectories after MLPG are those that will be used to reconstruct the speech waveform. The results are presented in Fig.~\ref{fig:mte_convergence}. It is observed that the trajectory error is reduced consistently by the MGE criterion, and converges after about 15 epochs. The `jump' in the trajectory errors at the $11^{\mathrm{th}}$ iteration is expected: we increased the value of momentum at that point. This phenomenon has been reported in~\cite{hinton2010rbm}. Similar convergence properties were also observed in the MGE-BN-DNN system, and is consistent with that reported in our previous work~\cite{wu2015minimum}.

We next compared the objective error of the DNN and BN-DNN architectures with and without MGE training. The objective results are presented in Table~\ref{tab:summary_objective}.
Compared to the DNN system without MGE training, MGE-DNN reduces MCD and $F_{0}$ RMSE from 4.19 dB and 9.13 Hz to 4.12 dB and 8.93 Hz, respectively. In comparison with BN-DNN, both MCD and $F_{0}$ RMSE measures for MGE-BN-DNN are reduced from 4.00 dB and 8.90 Hz to 3.97 dB and 8.89 Hz, respectively. The distortion reduction for BN-DNN is less than for DNN, and we think this is because BN-DNN already includes contextual constraints via the stacked bottleneck features at the input, and that these already improve the output trajectories.
 
\begin{figure*}[!t]
\centering
\includegraphics[scale=0.52]{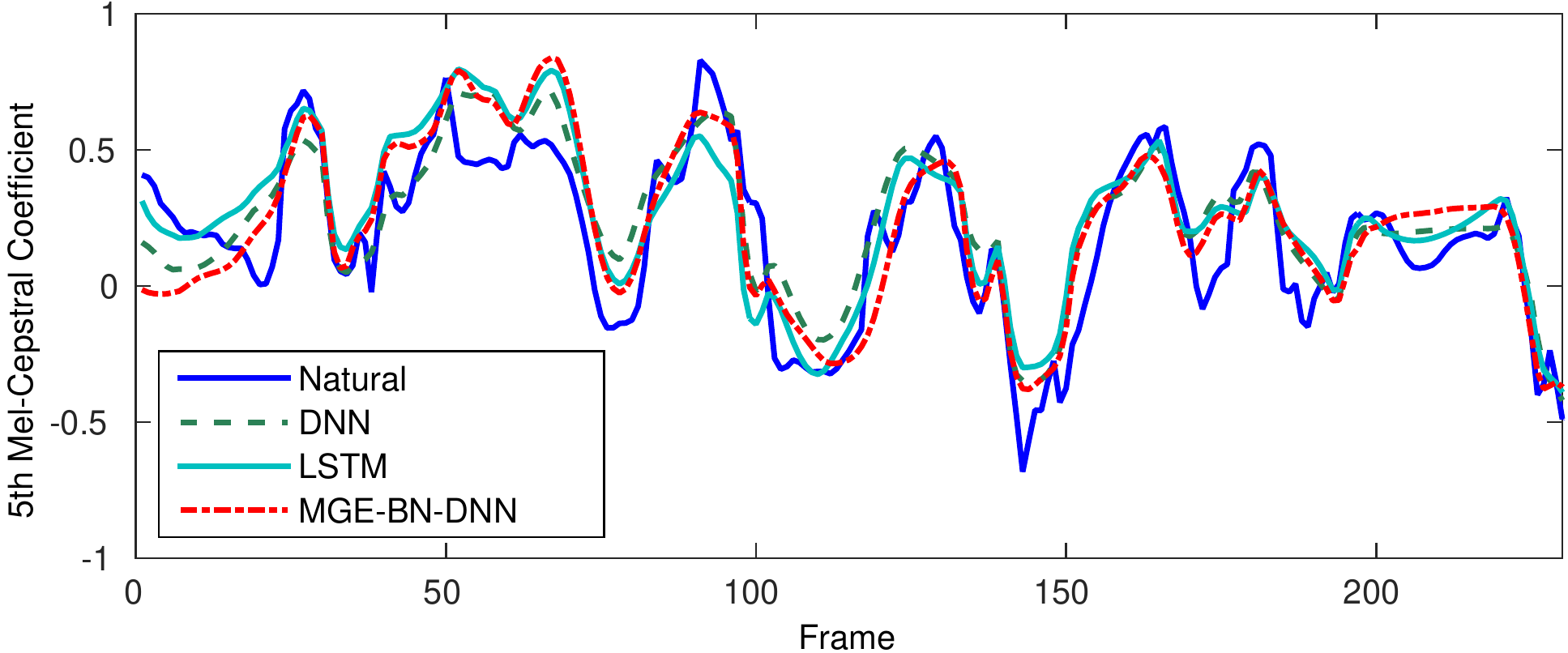} 
\vspace{-3mm}
\caption{Trajectories of the $5^{\mathrm{th}}$ Mel-Cepstral Coefficient (MCC) for one test utterance:  reference natural speech (Solid blue line); predicted by DNN (Dashed green line); LSTM (Solid light green line); proposed MGE-BN-DNN (Dashed red line).}
\label{fig:trajectory}
\end{figure*}

After that, we compared the performance of MGE-DNN and BN-DNN. In comparison with BN-DNN, MGE-DNN achieves higher distortions for all the measures. This indicates that stacking bottleneck features is more effective than the MGE criterion in improving the model accuracy. As we discussed above, MGE-BN-DNN can reduce the distortion further. This implies that even though MGE criterion alone is not as effective as stacking bottlenecks, it is complementary to stacking bottlenecks and the two techniques can be easily integrated to boost the performance. In summary, the objective results confirm the effectiveness of the proposed MGE criterion in improving DNN accuracy, and that MGE is complementary to the stacking of bottleneck features. Fig.~\ref{fig:trajectory} provides an example trajectory.

\subsubsection{Summary of objective evaluation}

Objective results for all systems on the evaluation set are presented in Table~\ref{tab:summary_objective}. The performance of all the systems was optimised on the development set.

Between the two baselines, LSTM is consistently better than DNN under all objective measures, consistent with~\cite{fan2014rnn}. MGE-DNN does not outperform LSTM objectively. The proposed MGE criterion only models the interaction between static and dynamic features, while the LSTM explicitly models temporal dependency in speech. 

All systems employing stacked bottlenecks and/or MGE training objectively outperform the DNN baseline. All systems employing stacked bottleneck features achieve lower MCD and BAP distortion than LSTM, although LSTM achieves slightly lower $F_{0}$ RMSE than BN-DNN and MGE-BN-DNN.

\subsection{Subjective evaluation}

Whilst objective measures are useful in system development, they are not always reliable predictors of listeners' preferences, so we also conducted a series of subjective preference tests. 30 paid native English speakers participated in each test, in which they each listened 20 randomly selected pairs of utterances and decided which item in each pair sounded more natural (or chose a ``no preference'' option). The utterances within each pair came from differing systems but had the same linguistic content.

\begin{figure}[!htb]
\centering
\includegraphics[scale=0.28]{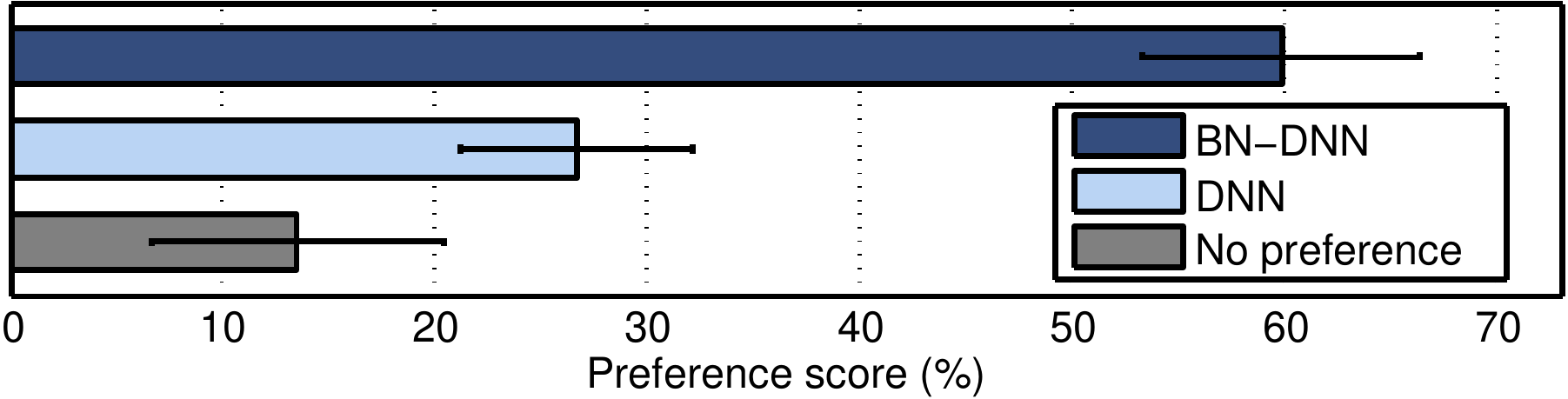} 
\vspace{-3mm}
\caption{Preference scores with 95\% confidence intervals for BN-DNN vs. DNN.}
\label{fig:bn-dnn_dnn}
\end{figure}

The preference scores for BN-DNN vs. DNN are presented in Fig.~\ref{fig:bn-dnn_dnn} and confirm that synthetic speech from the stacked bottleneck BN-DNN system is significantly more natural than DNN (60\% vs. 27\%). 

\begin{figure}[htb!]
    \centering
    \subfigure[BN-DNN vs BN-DNN-VB]{
        \includegraphics[scale=0.28]{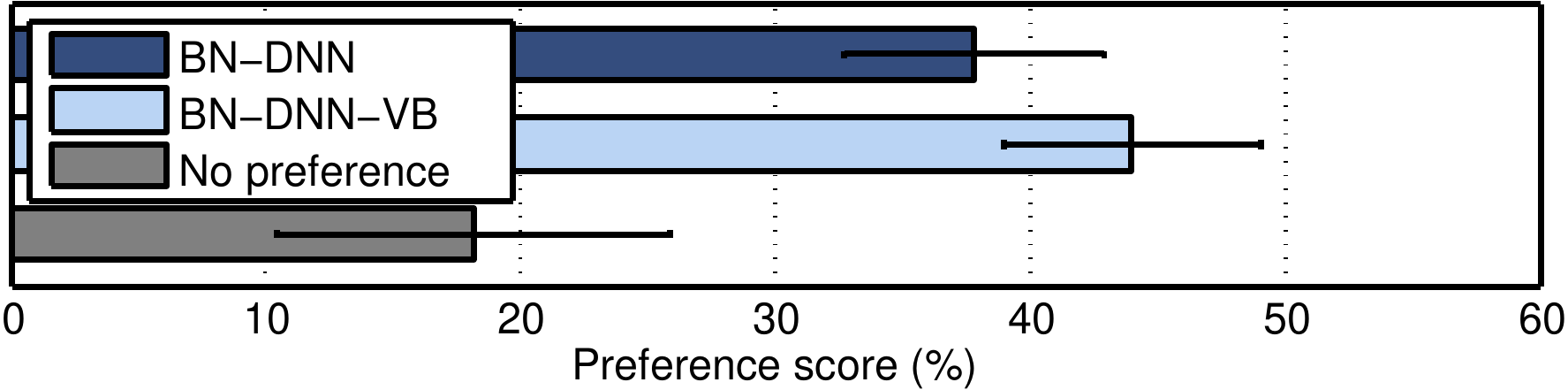}
        \label{fig:bn-dnn_bn-dnn-vb}
    }
    \subfigure[BN-DNN vs BN-DNN-MFC]{
        \includegraphics[scale=0.28]{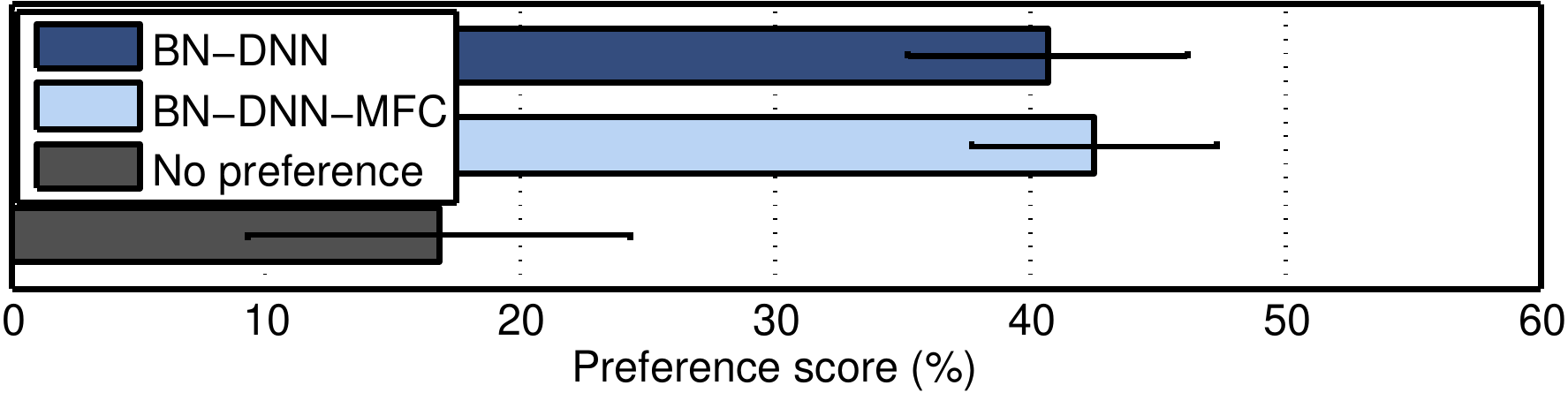}
        \label{fig:bn-dnn_bn-dnn-mfc}
    }
    \subfigure[BN-DNN vs BN-DNN-WSJ]{
        \includegraphics[scale=0.28]{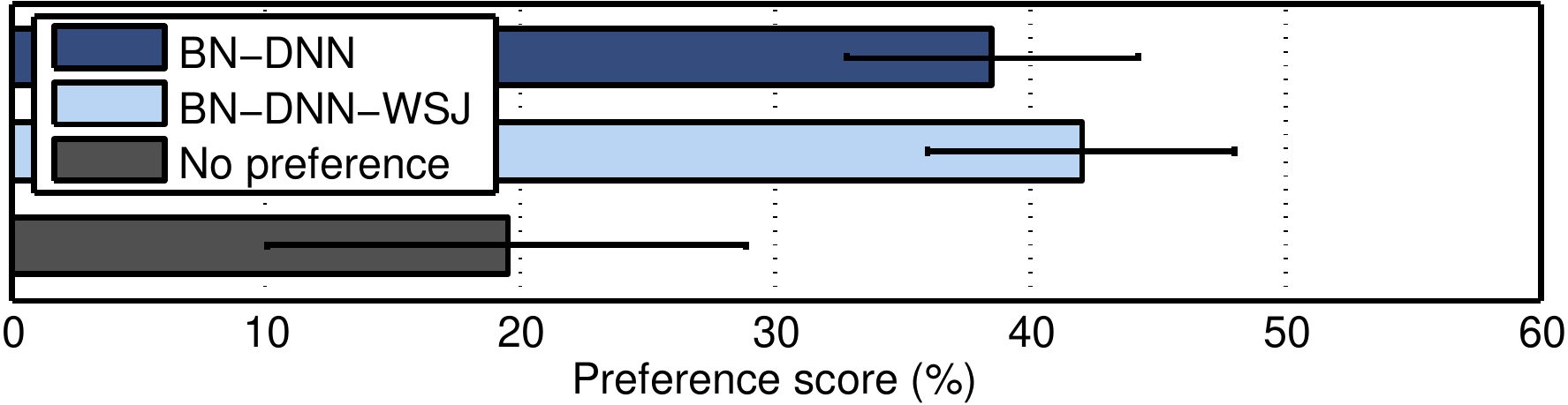}
        \label{fig:bn-dnn_bn-dnn-wsj}
    }
	\vspace{-3mm}
    \caption{Comparison between various ways to train the bottleneck network: Preference scores with 95\% confidence intervals. Refer to Section~\ref{sec:setups} for system descriptions.}
    \label{fig:bottleneck_features}
\end{figure}

Recall from Section~\ref{sec:setups} that there are several different ways to train the bottleneck network.  Fig.~\ref{fig:bottleneck_features} presents the results for the pairwise comparisons we made between these variants.  

The difference between BN-DNN-VB in Fig.~\ref{fig:bn-dnn_bn-dnn-vb} and BN-DNN-MFC in Fig.~\ref{fig:bn-dnn_bn-dnn-mfc} is in the acoustic feature representations and sampling rate (refer to Section~\ref{sec:setups} for details) and the results suggest that we can equally well use lower quality data and mismatched acoustic features to train the bottleneck network. Fig.~\ref{fig:bn-dnn_bn-dnn-wsj} shows that even out-of-domain data (a North American English speech recognition database) is effective for training the bottleneck network. There is an overall tendency that using more data (which is always also from multiple speakers) to train the bottleneck network is helpful, although the improvements over BN-DNN are not statistically significant.

That the data used to train the bottleneck network does not need to be speaker- or accent-specific is encouraging. Whether it needs to be language-specific is a question left for future experimentation.

\begin{figure}[!htb]
\centering
\includegraphics[scale=0.28]{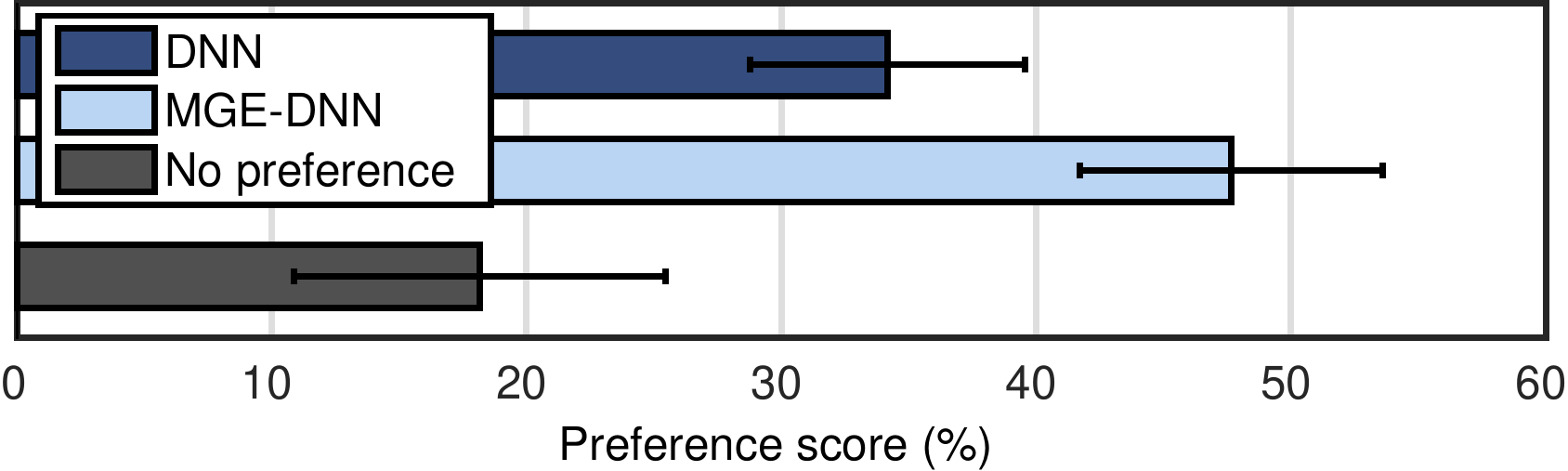} 
\vspace{-3mm}
\caption{Preference scores with 95\% confidence intervals for DNN vs. MGE-DNN.}
\label{fig:dnn_mte-dnn}
\end{figure}

The performance of the proposed MGE criterion was examined first in comparison to the baseline: preference scores for MGE-DNN vs. DNN are presented in Fig.~\ref{fig:dnn_mte-dnn} and show that MGE-DNN is significantly better than DNN in terms of naturalness.

\begin{figure}[!htb]
\centering
\includegraphics[scale=0.28]{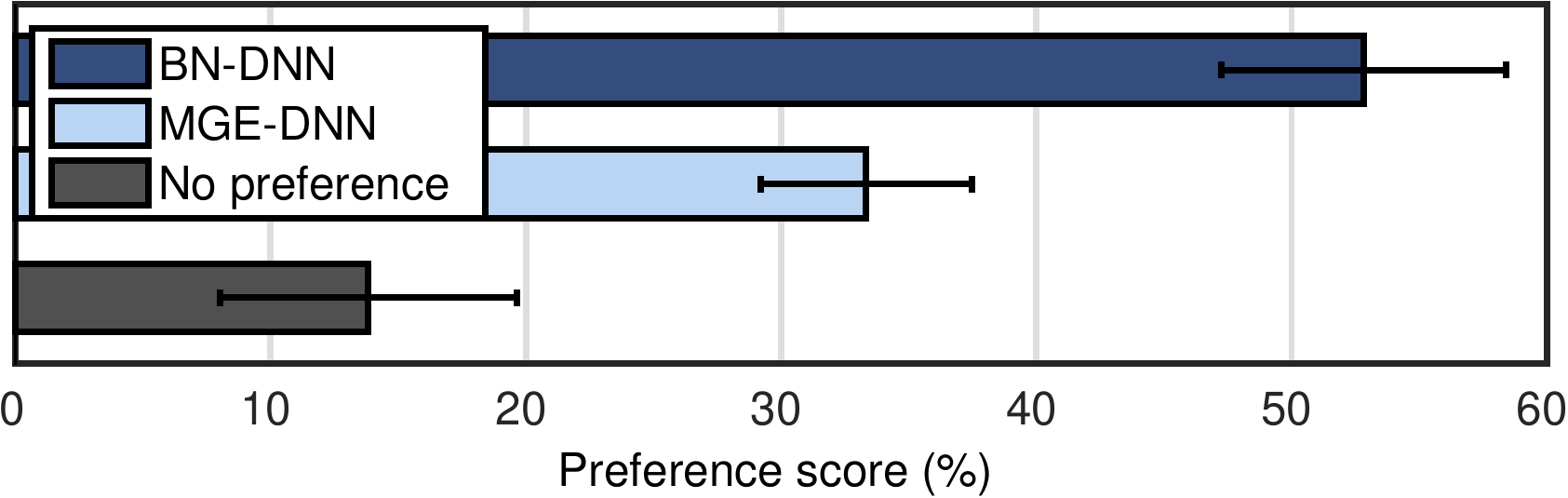} 
\vspace{-3mm}
\caption{Preference results between BN-DNN and MGE-DNN with 95\% confidence intervals.}
\label{fig:bn-dnn_mte-dnn}
\end{figure}

Next, we compared the performance of the two techniques proposed in this paper. In BN-DNN, contextual constraints are included at the input by stacking bottleneck features, while in MGE-DNN, contextual constraints are considered at the output by explicitly modelling the relationship between static and dynamic feature via the MGE training criterion. The preference scores in Fig.~\ref{fig:bn-dnn_mte-dnn} show that BN-DNN is very substantially (and statistically significantly) better than MGE-DNN: stacking bottleneck features is much more effective than MGE training.

\begin{figure}[!htb]
\centering
\includegraphics[scale=0.28]{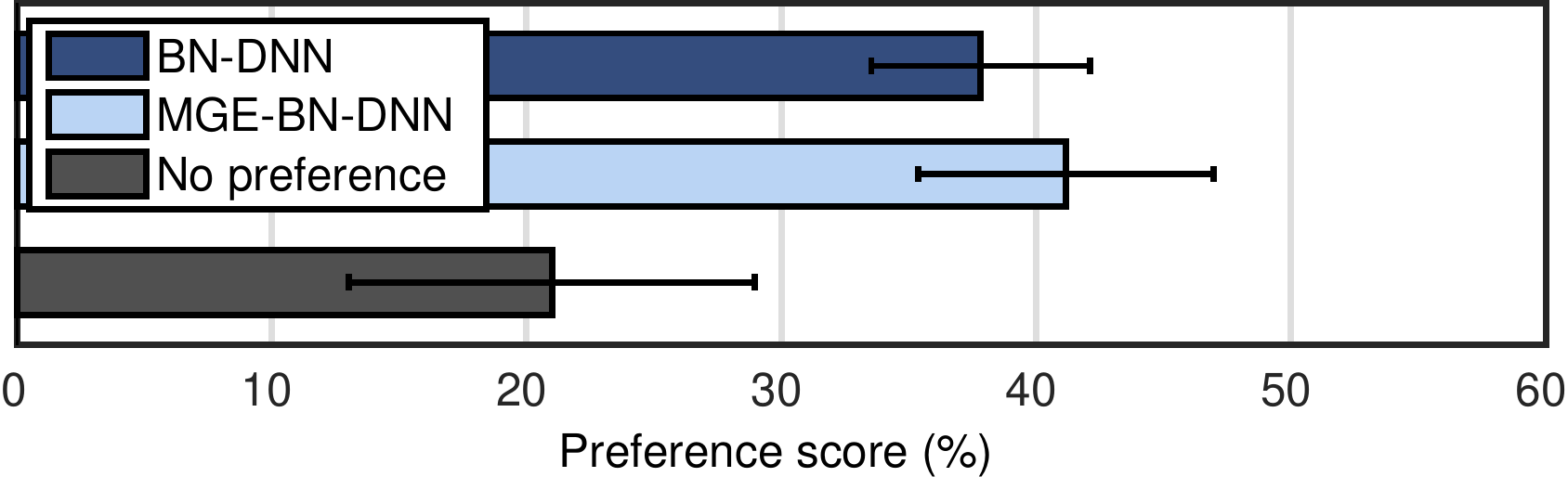} 
\vspace{-3mm}
\caption{Preference scores with 95\% confidence intervals for BN-DNN vs. MGE-BN-DNN.}
\label{fig:bn-dnn_mte-bn-dnn}
\end{figure}

Of course, the two techniques can be combined, and the preference scores for BN-DNN vs. MGE-BN-DNN are presented in Fig.~\ref{fig:bn-dnn_mte-bn-dnn}. These indicate that MGE may further improve naturalness on top of the improvements already obtained by stacking bottleneck features, although the difference is not significant.  Overall, these preference tests demonstrate the effectiveness of the proposed MGE criterion, and reconfirm our findings reported earlier in~\cite{wu2015minimum}.

\begin{figure}[!htb]
\centering
\includegraphics[scale=0.28]{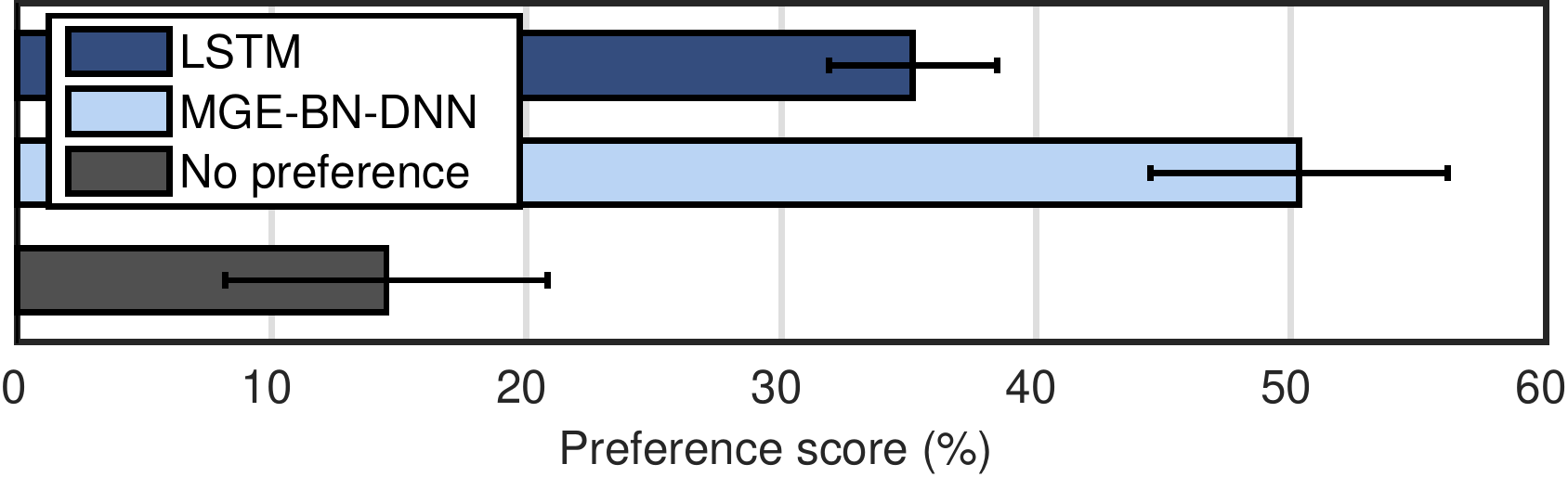} 
\vspace{-3mm}
\caption{Convergence of the proposed minimum generation error training criterion for the DNN system (MGE-DNN).}
\label{fig:mte-bn-dnn_lstm}
\end{figure}

Finally, we compared the combined proposed techniques (MGE-BN-DNN) to the baseline LSTM system. Fig.~\ref{fig:mte-bn-dnn_lstm} shows that in our setting MGE-BN-DNN is significantly better than LSTM in terms of naturalness (50\% vs. 35\%).

One interesting observation is that the preference scores just presented are in fact consistent with the objective results presented earlier (especially MCD), despite our own feeling (widely shared across the community) that objective measures do not reliably correlate with human judgements.

\section{Discussion}

Our experimental results confirm the effectiveness of both proposed techniques as ways to include contextual constraints in neural network-based speech synthesis. Although LSTM-based recurrent neural networks may provide an elegant way to include temporal constraints, there are at least two reasons to choose our proposed framework of combining stacked bottleneck features with MGE training.

First, the proposed framework can make use of out-of-domain data in a very straightforward way: the bottleneck network can be trained using relatively low-quality speech, from multiple speakers. Since thousands of hours of speech data are available for training speech recognition systems, these data might also be used to train the bottleneck network. Using such large amounts of data might further improve the results presented here, where we used tens of hours data to train the bottleneck network.

Second, although the total training time of the proposed framework is close to that of the LSTM system, at synthesis time the computational complexity of the proposed framework is considerably lower. In particular, to generate 142 utterances (all the utterances in the development and evaluation sets), the LSTM system took 215 seconds\footnote{Times reported are for neural network computations only and do not include lingusitic feature extraction or vocoding to reconstruct speech waveform. The hardware used here is Nvidia GTX TITAN with 2688 cores 6G RAM.}, while the proposed MGE-BN-DNN system took only 8 seconds. The computational cost of the LSTM system is about 27 times higher than the proposed framework, during synthesis.

\section{Conclusion}

We propose two techniques to improve the performance of DNN-based speech synthesis, namely stacked bottlenecks and a minimum generation error training criterion. The two techniques can be easily combined in a single DNN speech synthesis framework. This novel framework allows us a) to benefit from additional out-of-domain data to improve the synthesis performance; and b) to include contextual constraints without much increase in computational complexity at synthesis time. Both objective and subjective results confirm the effectiveness of the proposed system over both DNN and LSTM baselines.

\noindent To summarise the main findings:
\begin{itemize}
	\item Stacking bottleneck features provides an effective way to include contextual constraints. As shown in the experiments, by setting the size of the bottleneck layer to 32, we can effectively include contextual constraints from 23 consecutive frames, which span a segment of $23 \times 5 = 115$ ms. Because the dimensionality is low, the computational cost of the synthesis network does not increase significantly.

	\item Out-of-domain data (e.g., lower quality data collected for speech recognition) can be used to train the bottleneck network. This provides a flexible way to benefit from the readily-available large quantities of such data.

	\item The minimum generation error training criterion is effective and can improve model accuracy, as shown in the experiments. As the criterion is only employed at the training stage, it does not introduce any additional computational complexity at synthesis time.
	
	\item The two techniques, stacked bottleneck features and minimum generation error training criterion, are complementary. The techniques are applied at the input and output, respectively, and can be easily combined.
\end{itemize}

Currently, the bottleneck network and the synthesis network are trained independently. The performance might be boosted if the two networks were to be optimised jointly. Combining stacked bottleneck features with an LSTM-based synthesis network might also be effective. Our preliminary results show that the proposed bottleneck features can also be used to guide rich-context model selection~\cite{merrit2015rich}, and waveform unit selection~\cite{merrit2016hybrid}. We will leave those directions of research as follow-up work for the future. 

%The structure of sentences may also have impact on the effectiveness of the  trajectory modelling method.
% \fi
%\IEEEpeerreviewmaketitle

%\vspace{1mm}
%\scriptsize
%\textbf{Acknowledgements:} this work was partially supported by EPSRC under Programme Grant EP/I031022/1 (Natural Speech Technology).

\ifCLASSOPTIONcaptionsoff
  \newpage
\fi



%\bibliographystyle{IEEEtran}
\bibliographystyle{IEEEtran}
\bibliography{refs.bib}


\end{document}